\documentclass{aa}

\usepackage{graphicx}
\usepackage{epsfig}

\begin{document}

   \title{HST-1 as a Window to the Energetics of the Jet Spine of M\,87}

   \author{B. Punsly}

   \institute{1415 Granvia Altamira, Palos Verdes Estates CA, USA
90274: ICRANet, Piazza della Repubblica 10 Pescara 65100, Italy and
ICRA, Physics Department, University La Sapienza, Roma,
Italy\\
\email{brian.punsly@cox.net}\\ }

   \date{Received May 7, 2023; }
\titlerunning{HST-1: a Window to the Jet Spine of M\,87}

\abstract {A new interpretation of the optical knot in the jet of M\,87, HST-1, is presented. High sensitivity 22 GHz Very Large Array images locate HST-1 to within 6 mas of the jet axis immediately upstream. 1.7 GHz Very Long Baseline Array images of a bright flare in 2005 indicates that the preponderance of emission in the early stages originates in an elongated region that is tilted $12.5^{\circ}$ from the jet axis. The superluminal motion, shape, location and the large jet-aligned optical/UV polarization suggest an identification with the putative relativistic spine of the jet. As such, energy flux estimates for HST-1, $\sim 870$ mas from the nucleus, published in 2006 indicate that the central engine injected $Q_{\rm{spine}}\approx 2.5 \times 10^{41}\rm{ergs/s}$ into the base of the spine $\sim 200$ years earler. Furthermore, previous studies reveal a tubular protonic jet on sub-mas scales that envelopes a low luminosity core, presumably the faint spine base. It was estimated that the central engine injected $Q_{\rm{tubular\,jet}}\approx 6.1\times 10^{41}\rm{ergs/s}$ $\sim 1.5$ years earlier. If one component of the jet is inherently more powerful, a firm constraint on total jet power in the recent past exists. If the emitted jet is inherently dominated by the spine (tubular jet) then the total bilaterally symmetric jet power emitted from the central engine was $<4Q_{\rm{spine}}\approx 1.0 \times 10^{42}\rm{ergs/s}$ ($< 4Q_{\rm{tubular\,jet}}\approx 2.4\times 10^{42}\rm{ergs/s}$) $\sim 200$ ($\sim 1.5$) years earlier. Assuming a nearly constant central engine injected jet power for $\sim 200$ years indicates a total jet power of $\lesssim  2\times 10^{42}$ ergs/s in epochs of modern observation or $\lesssim 3.5\%$ jet production efficiency for an accretion rate of 0.001$M_{\odot}$/yr. Seemingly, the focus of Event Horizon Telescope Collaboration numerical models should be biased towards jet powers $\lesssim 2 \times 10^{42}$ ergs/s as opposed to larger estimates from ejections many centuries or millennia earlier.}


   \maketitle
%

\section{Introduction}
The nearby galaxy M\,87 ($\approx 16.8$ Mpc distant) possesses the most studied astrophysical jet. A bright optical knot was found with the Hubble Space Telescope (HST) $\sim 870$ mas from the nucleus \citep{bir99}. It was the most superluminal feature ever witnessed in the jet which was typified by subluminal motion at that time. Thus, it experienced tremendous observational attention. In early 2005, it flared in the optical/UV and X-ray to be brighter than the nucleus \citep{har06}. It was a likely candidate to be the source of a TeV flare in this epoch \citep{abr12}. HST-1 has been described in terms of a re-collimation shock \citep{sta06} This article provides an alternative explanation as a dissipative region of the relativistic central spine of the jet.

\par The notion of a sheath and a highly relativistic spine has been used to understand high energy phenomena in blazars and M87, in particular, that were difficult to reconcile with single zone models \citep{ghi05,tav08}. However, direct observation of the physical nature and dynamics of the spine has been elusive. On sub-mas scales, cross-sections of the jet in the highest sensitivity images (as of 2022) with 43 GHz and 86 GHz Very Long Baseline Interferometry (VLBI), in 2013 and 2014 respectively, detect a predominantly double-ridged (edge brightened) morphology \citep{wal18,had17,pun23}. Analysis of the large (ridge) peak to central trough intensity ratios in cross-sectional slices require a source that is a bright thick-walled tubular jet that envelopes a nearly invisible core or spine at  $0.35 \,\rm{mas}<z< 0.65 \,\rm{mas}$, where $z$ is the axial displacement from the nucleus \citep{pun23}. New high resolution, 86 GHz VLBI with the Global Millimetre VLBI Array, the phased Atacama Large Millimetre/submillimetre Array and the Greenland Telescope in 2018 have detected a very narrow central feature that points back towards the vicinity of the ring of emission surrounding the black hole \citep{lu23}. Curiously, the detected feature is bright enough that it would have been detected in the 2013 and 2014 cross-sections (even with lower resolution), but was not apparent \citep{pun22,pun23}. It would seem to a be variable feature that is not modeled in the numerical simulation library of the Event Horizon Telescope Collaboration, EHTC hereafter \citep{por19}. An order of magnitude farther out there is observational evidence of faint disjoint narrow features along the axis, but nothing that connects back close to the source \citep{asa16,had17}. The dynamics of the spine within an arc-second of the source is unknown. This motivates efforts to find direct observational evidence of strong spine dissipation that might reveal its energetics and composition.

\par Previously published VLBI images that clearly resolve the jet from HST-1 in the axial direction have not been sensitive enough to detect jet emission that extends continuously to HST-1. There is a detection gap, from $z\sim 400$ mas to HST-1, $z\sim$ 870 mas, even with low frequency Very Long Baseline Array (VLBA) observations, 1.7 GHz and 327 MHz \citep{che07,ram09}. In Section 2, new sensitive high dynamic range images from archival Very Large Array (VLA) data at 22 GHz are able to resolve HST-1 from the jet in the axial direction and trace the jet direction continuously in the gap. There is a mild bend of the jet northward, so that HST-1 lies along the jet centerline or ``spine". It is argued in Section 3 that the flare in HST-1 that began in 2005 might be a ``spinal disruption event". It is ideal for the purposes of determining energetics. There are a wealth of observations over many epochs and frequency \citep{per11,har06,har09}.\footnote{A line of sight to the jet (LOS) of $18^{\circ}$ is chosen throughout.}
\par In Section 4, previous equipartition models of the flare in 2005 from \citet{har06} are interpreted in the context of the spine. In Section 5, estimates of the energy fluxes of the spine and the surrounding tubular jet from sub-mas scales are combined to make a unified picture of the jet over the last 200 years. The jet power in recent epochs is crucial to the EHTC data analysis. The data is difficult to interpret and it has become conflated with a library of numerical simulations which in turn are down-selected using the constraint of a viable jet power. A better knowledge of the jet power is more important than ever as the conflation process is becoming more aggressive. It is now proposed that image reconstruction should be biased by ``viable" numerical models \citep{med23}. The range of viable jet powers considered by the EHTC is a few hundred and the result here is at the very low end of the range.

\begin{figure*}
\begin{center}
\vspace{-2cm}
\includegraphics[width=160mm, angle=0]{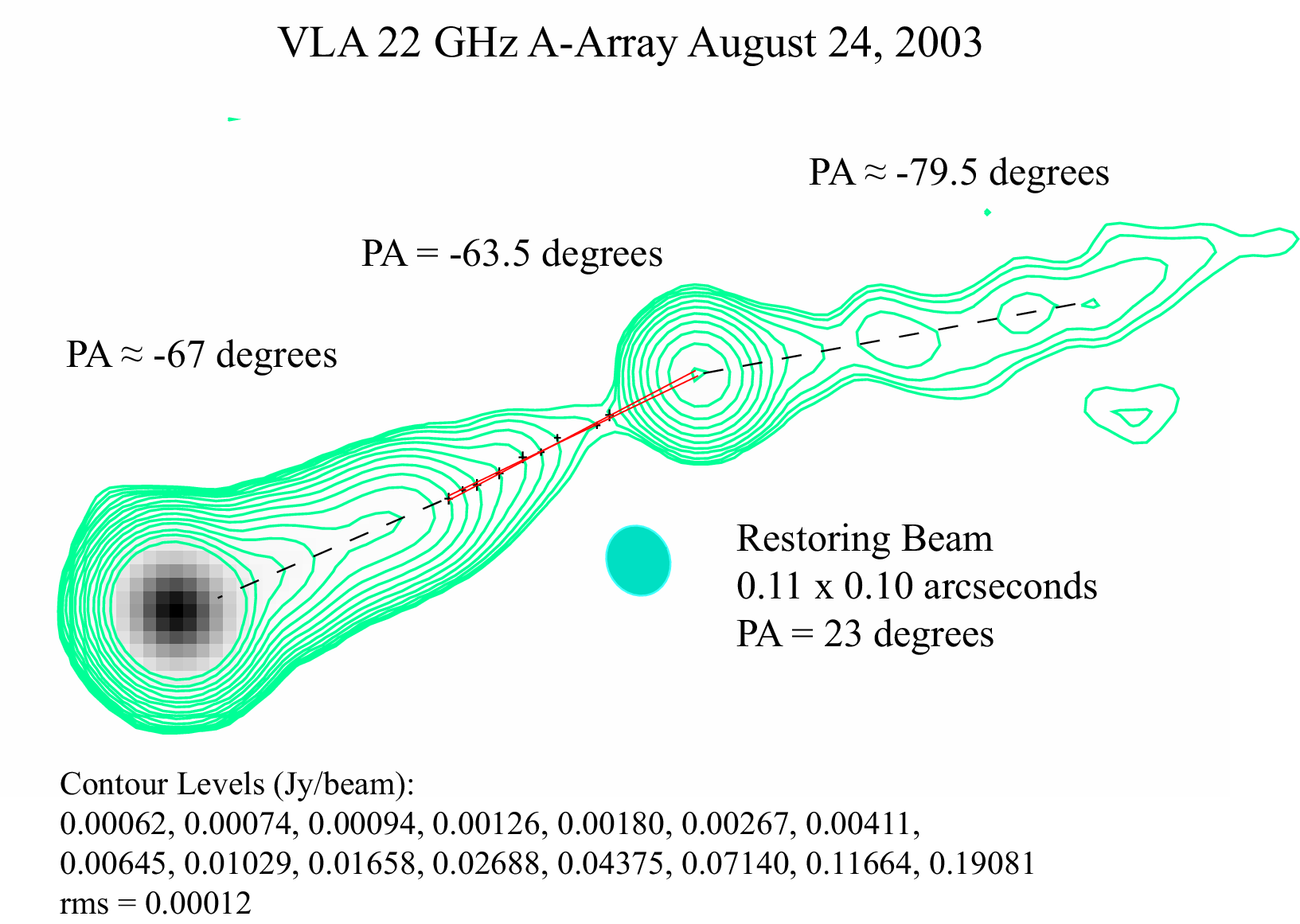}
\includegraphics[width=160mm, angle=0]{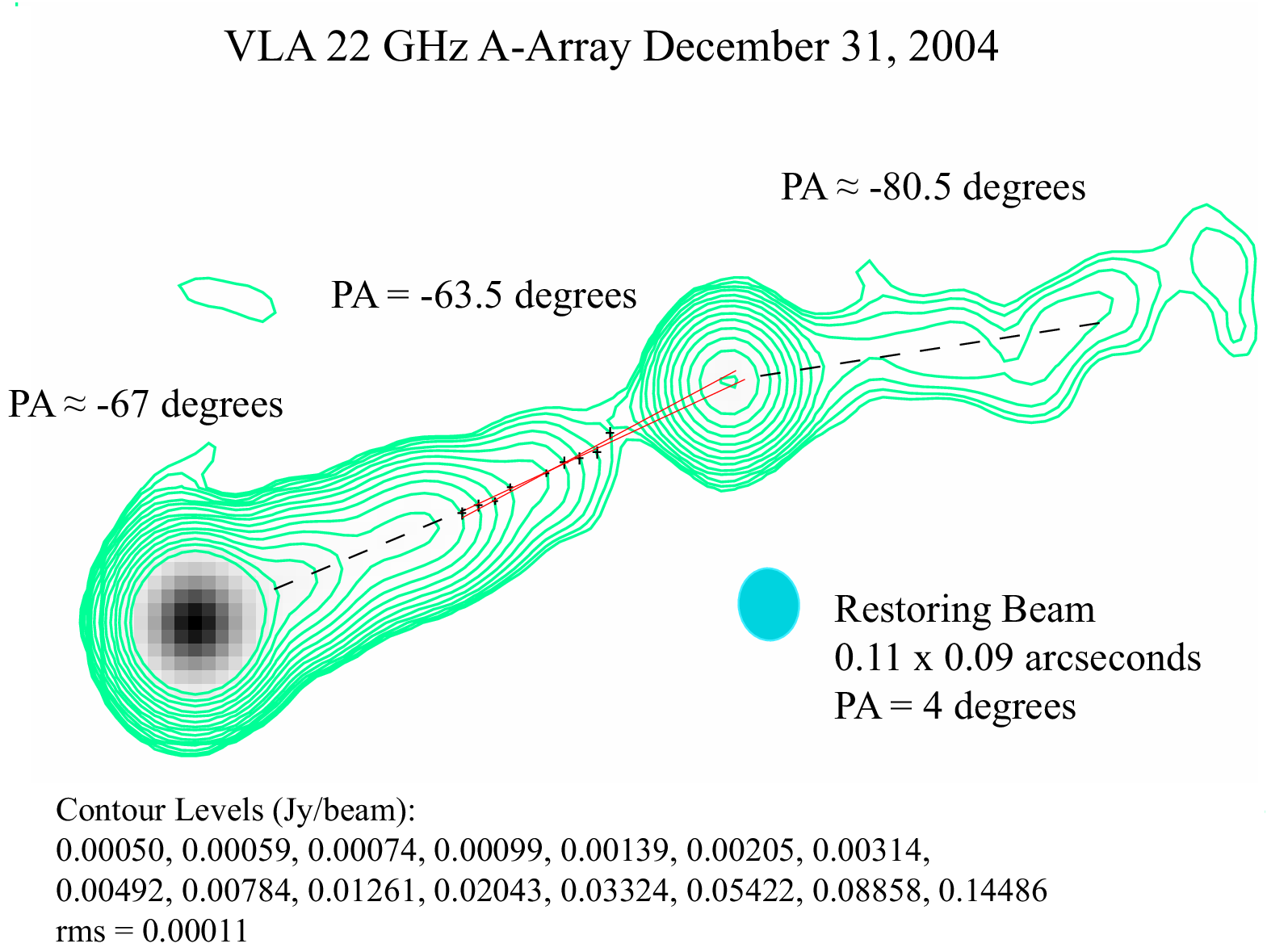}
\caption{\small{The 22 GHz images have the sensitivity to reveal the jet direction in the gap (in 1.7 GHz and 327 MHz VLBA images) between $z \approx 400$ mas and HST-1. The red lines in the gap are the range of the standard error in the linear fit to the peak intensity (identified with the jet center-line). HST-1 lies on a short extrapolation of the jet center-line to within $\pm6$ mas. The image from December 31, 2004 is at the beginning of the flare. The Gaussian beam FWHM in blue provides a scale for the image.}}
\end{center}
\end{figure*}

\section{Imaging the Adjacent Jet Upstream of HST-1}
The 1.7 GHz VLBA observations are not sensitive enough to detect emission in the gap 400 mas $<z< 870$ mas \citep{che07}. Thus, they cannot determine the local jet direction just upstream of HST-1. The best observations for imaging this region are 22 GHz VLA due to the sensitivity. The synthesized beam is about 2-3 times the jet width based on extrapolating the outer edges of the 1.7 GHz VLBA jet in Figure 1 of \citet{che07}. Some 22 GHz images were previously published \citep{che11}. Yongjun Chen graciously recreated the image FITS files for two of these observations with improved signal to noise for the purposes of this article. Inspection of the residual images associated with the 2011 paper, showed a significant emission structure pattern that was morphologically similar to the original structure. This means there was still some emission left in the residual image. This suggests that more ``cleaning" was required to create a new residual image that looks like the noise distribution over the whole field, thereby improving the signal to noise ratio. An earlier version of the 12/31/2004 image appeared in \citet{che11}. In spite of the modest resolution, intensity cross-sections orthogonal to the jet determine the centroid of the local jet emission to $\approx 1/10$ of the synthesized beam width for a signal to noise ratio $>5$ \citep{con98}. The centroids of the cross-sections define the jet direction. This is achieved by linear fitting these centroids over the range, $z\sim 400$ mas to $z\gtrsim 700$ mas, using least squares with uncertainty in both variables \citep{ree89}. Since the beam is considerably wider than the jet, the peak intensity will represent the centroid position. The uncertainty of the peak position is 1/10 the synthesized beam, unless the maximum is achieved at more than one point. The cross-section can be tangent to the contour over a finite range of points which can be noticed with large magnification of the image. In this circumstance, the uncertainty is the distance between the maxima added in quadrature with 1/10 of the synthesized beam. Finding the position angle (PA) of the parallel cross-cuts is an iterative process. The intensity peaks are found and a ``line of centroids" fitted. The cross-cut PA is varied until the fitted line of centroids is perpendicular to the cross-cuts. Figure 1 shows two epochs in order to see if the jet direction found is independent of observation. The contours were chosen to be approximately evenly spaced in the $z$ coordinate (approximately 1/4 to 1/3 of a beam width in spacing). Each fit has the same number of points. The process is not completely uniform or perfect. There is always one spacing per fit that is approximately twice as large as the others when using the log scale option for the contour spacing. The points that are chosen for the fit lay on the extremum of the $z$ coordinate of the contour and correspond to the location where the cross-cut is tangent to the contour. Since, the fit and the radio image are created with two different softwares, this choice of points facilitates a very accurate overlay (alignment) of the fit on the radio image. Thereby, providing the reader with an accurate fitted jet axis with its uncertainty on the radio image itself.  The range of fits (the red lines) is indicated by the standard error of the regression \citep{ree89}. 2003 has a smaller spread in the standard error of the fit because HST-1 was fainter and it did not skew the centroid of the flux density beyond 700 mas (i.e., there is an additional, closer, reliable centroid in the linear fit). The standard error from the best fit line is maximal at the endpoints of the fitted region, reaching $\sim \pm 3$ mas. Based on a short extrapolation of the fit, the primary result is that the position of HST-1 aligns with the jet axis that is immediately upstream to within $<6\,\rm{mas}$ in 2003.
\par Figure 1 traces the jet trajectory with 3 linear pieces. The inner jet, $z< 400 \rm{mas}$, is indicated by a black dashed line with the traditional  $\rm{PA}=-67^{\circ}$ \citep{had16}. The outer dashed black line is an ``eyeball" fit to the faint trajectory beyond HST-1. It is significant that the jet which appears to be very straight for 870 mas makes an abrupt bend at HST-1 that is illustrated in Figure 1. This occurs before the flare and persists during the flare. This would need to be an integral part of any physical description of HST-1.

\begin{figure}
\begin{center}
\includegraphics[width=75mm, angle=0]{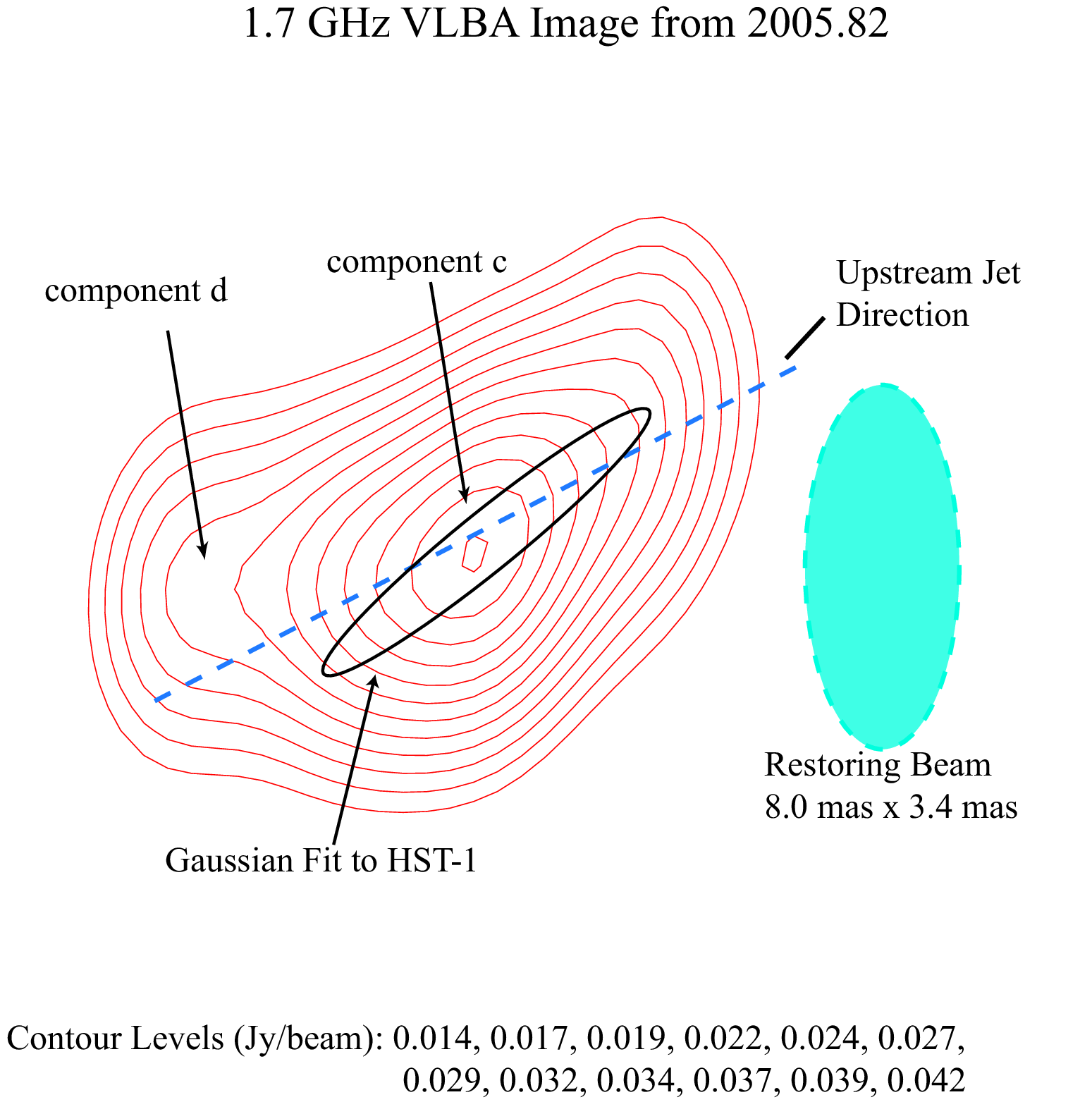}
\caption{The 1.7 GHz VLBA image from 2005.82 of HST-1. The Gaussian fit to the component ejected during the 2005 flare, component c, is very elongated along a direction that is close to the local upstream jet direction in Figure 1. The single Gaussian fit is an ellipse, 9.2 mas x 1.3 mas (unresolved) at $\rm{PA} = -51^{\circ}$. Components a and b are off to the right and are too faint to be revealed by this contour map. The Gaussian beam FWHM in blue provides a scale for the image.}
\end{center}
\end{figure}

\section{Is it Plausible to Associate HST-1 with the Jet Spine?}
\par Figure 2 is  a 1.7 GHz image from 2005.82 of HST-1 with a Gaussian fit (both the image FITS file and Gaussian fit were generously provided by C.C. Cheung). The fit and the image are for the dominant, component c that seems to be responsible for the flare in 2005. The fit does not pertain to the weaker components. It is the same image FITS file used to create the insert in the right hand corner of Figure 1 of \citep{che07}. All the components were briefly described as unresolved \citep{che07}. No Gaussian fits were published, but an elliptical fit to component c was made in support of the analysis of \citep{che07}. It is not a fit to the entire HST-1 complex. Adding a weak point source at the eastern edge (component d) \textbf{might} make a small change. The components a and b are too faint to appear given the lowest contour level that is chosen in the image of Figure 2 of this paper. At this early stage of flare evolution, the Gaussian fit is a very elongated feature that is formally unresolved, transversely. Note, it is the best fit assuming just one component that makes up the preponderance of flux emitted by the flare. The main conclusion of this fit is that it is necessarily highly elongated and it is also rather closely aligned with the local jet PA found in Figure 1, $\Delta\rm{PA}=12.5^{\circ}$.

\par The putative spine is a powerful, highly relativistic central component of the jet. It is generally believed to be a Poynting jet with an ordered magnetic field that is predominantly toroidal \citep{ghi05,gab18}. At high frequency (optical and UV), the Faraday rotation that is prominent at radio frequencies is minimal since the rotation angle scales inversely with the frequency squared \citep{che11}. Consequently, the observed polarization direction should represent the intrinsic polarization of the emitted radiation at its source. Thus, one expects very large optical and UV polarization aligned with the jet direction when the spine radiates \citep{gab19}.

\par By comparison, in 2005, HST-1 is a historically bright superluminal knot in the M87 from P-band to X-rays. It is located to within 6 mas of the jet center-line. The initial configuration of the luminous ejection in 2005 is very elongated and almost parallel to the local upstream jet axis. It is noted that the ejections in the complex as well as the direction of the parsec scale jet change direction over time \citep{gir12,ro23}. However, the initial elongation of the flaring knot, HST-1, is closely aligned with the jet direction. It has high (25\%-40\%) optical and UV polarization that is a aligned within a few degrees of the local jet axis \citep{per11}. This is either a group of strong coincidences or the 2005 flare of HST-1 arises from dissipation of the spine. It is the primary tenet of this paper that these are too many coincidences to ignore.

\section{Energy Flux Estimates of HST-1 in 2005}
In this section, the previous equipartition model of HST-1 in \citet{har06} is revisited in the context of the energetics of the spine. One of the big unknowns was the apparent velocity of the ejected material that formed the predominant contribution to the luminosity of HST-1 in 2005. It was subsequently shown with 1.7 GHz VLBA that a powerful knot of emission emerged in 2005.04 the component c shown in Figure 2) and traveled downstream at $\beta_{\rm{app}}=v_{\rm{app}}/c =1.14\pm 0.14$ \citep{che07}. $\beta_{\rm{app}}$ constrains the range of viable models in \citet{har06}.
\begin{table*}
\caption{Parameters of Harris et al. (2006) Equipartition Models}              
\label{table:1}      
\centering                                      
\tiny{\begin{tabular}{c c c c c c c c c c c c}          
\hline\hline                        
(1) & (2) & (3) & (4)  & (5) & (6) & (7) & (8) & (9) & (10) & (11)\\
$\delta$ & $B$ & Radius & $E_{\rm{min}}$ & $Q_{\rm{fill}}$ & $\beta$ &$\Gamma$ & $E^{\perp}$ & Poynting Power &Kinetic Power& $Q_{\rm{total}}$\\
      & (G) & (cm) & (ergs)& (ergs/s)&(c)& & (statvolt/cm)& (ergs/s) & (ergs/s)& (ergs/s)\\
\hline                                   
    1.0 & 0.10 & $8.3 \times 10^{16}$ & $5.2 \times 10^{48}$ &$1.2 \times 10^{42}$ & \textbf{0} &\textbf{1} &\textbf{0} & \textbf{0}&\textbf{0} &\textbf{0}\\
    \textbf{2.5} & \textbf{0.016} & $\mathbf{2.1 \times 10^{17}}$ & $\mathbf{1.7 \times 10^{48}}$ & $\mathbf{1.6 \times 10^{41}}$ & \textbf{0.79}& \textbf{1.6}
    &\textbf{0.02} & $\mathbf{1.6\times 10^{41}}$ & $\mathbf{8.0\times 10^{40}}$ & $\mathbf{2.4\times 10^{41}}$\\
    \textbf{2.6} & \textbf{0.014} & $\mathbf{2.2\times 10^{17}}$ & $\mathbf{1.6 \times 10^{48}}$ & $\mathbf{1.4 \times 10^{41}}$ & \textbf{0.82}& \textbf{1.7} &\textbf{0.02} & $\mathbf{1.7\times 10^{41}}$ & $\mathbf{8.4\times 10^{40}}$ & $\mathbf{2.5\times 10^{41}}$\\
    \textbf{2.7} & \textbf{0.014} & $\mathbf{2.2\times 10^{17}}$ & $\mathbf{1.5 \times 10^{48}}$ & $\mathbf{1.2 \times 10^{41}}$ & \textbf{0.84}& \textbf{1.8} &\textbf{0.02} & $\mathbf{1.7\times 10^{41}}$ & $\mathbf{8.7\times 10^{40}}$ & $\mathbf{2.6\times 10^{41}}$\\
    3.0 & 0.010 & $2.5 \times 10^{17}$ & $1.3 \times 10^{48}$ &$1.0 \times 10^{41}$ & \textbf{0.90}& \textbf{2.3} & \textbf{0.02} & $\mathbf{2.2\times 10^{41}}$&$\mathbf{1.1\times 10^{41}}$ & $\mathbf{3.3\times 10^{41}}$\\
\hline                                             
\end{tabular}}
\end{table*}
From \citet{ree66} and \citet{gin69},
\begin{equation}
\beta_{\rm{app}} \equiv \frac{v_{\rm{app}}}{c} = \frac{\beta \sin{\theta}}{1-\beta\cos{\theta}}\;,
\end{equation}
where $\theta$ is the LOS (chosen to be $18^{\circ}$ here) and $\beta$ is the velocity of HST-1 viewed in the cosmological rest frame of M\,87 with $\theta =90^{\circ}$ (M87 frame hereafter). Table 1 is arranged as follows. Everything in bold face is new and the other is data from \citet{har06}. Equation (1) indicates a narrow range of Doppler factors, $\delta$, in column (1) associated with $\beta_{\rm{app}}=v_{\rm{app}}/c =1.14\pm 0.14$ (in boldface), where
\begin{equation}
 \delta = 1/[\Gamma (1- \beta\cos{\theta})]\;.
 \end{equation}
 The next three columns are the magnetic field, $B$, the plasmoid radius from time variability arguments and the energy stored in the plasmoid, respectively from the models described in \citet{har03,har06}. In column (5), $Q_{\rm{fill}} = E_{\rm{min}}/T$, is the energy flux required to fill the plasmoid in the variability time scale, $T$ \citep{har06}. $T$ was estimated by \citet{har06} from the 2005 X-ray light curve.. The next two columns are the velocity and Lorentz factor in the M87 frame.
\par The remaining columns are new. They are motivated by the fact that the spine is believed to be a Poynting jet where $B$ in column (2) is an ordered magnetic field. From angular momentum conservation, $B$ is almost purely toroidal and $B^{\phi} \approx \Gamma B$ in the M87 frame \citep{pun08}. The poloidal Poynting power, $S^{z}$, in perfect MHD and approximate azimuthal symmetry is \citep{pun08},
\begin{eqnarray}
& S^{z} = \frac{\rm{c}}{4\pi}\int{-B^{\phi}E^{\perp}dA_{\perp}}\;,\quad E^{\perp}\approx -\beta B^{\phi} \;,\nonumber\\
& S^{z} =2\rm{c}\int{\Gamma^{2} \beta U_{B}dA_{\perp}}\;,
\end{eqnarray}
where ``$\perp$" is the orthogonal direction to $z$ and $\phi$ (azimuthal angle) and the normal cross-sectional area element is $dA_{\perp}$. $U_{B}$ and $U_{p}$ are the energy densities of the field and particles in the jet rest frame. $E^{\perp}$ and $S^{z}$ are tabulated in columns (8) and (9), respectively. Column (10) is the particle energy flux in the M87 frame, $\mathcal{K}$. The total jet power in column (11) is
\begin{equation}
Q_{\rm{total}} = S^{z} +\mathcal{K} \;,\quad \mathcal{K}=\int{U_{p}\Gamma^{2}\beta\rm{c}dA_{\perp}}\;, U_{p}=U_{B}\;.
\end{equation}
$Q_{\rm{total}}$ is larger than $Q_{\rm{fill}}$ in the range of $\delta$ relevant to HST-1 in 2005.
\section{Discussion and Conclusion} Section 2 showed that HST-1 lies along the central axis of the jet. In Section 3, it was argued that the superluminal motion, shape, location and the large, axis-aligned, optical/UV polarization strongly support an identification with the relativistic spine of the jet. In Section 4, the equipartition models in Table 1 indicate an energy flux $Q_{\rm{spine}}\equiv Q_{\rm{total}}\approx 2.5 \times 10^{41}$ ergs/sec. As mentioned in the Introduction, the spinal disruption analysis provides an alternative to re-collimation shock models of HST-1. As such, it is not intended to critique the interpretation that HST-1 might be a re-collimation shock.

\par A possible explanation of the sudden spine dissipation is given in Figure 1. The jet slowly drifts a few degrees for $z<870$ mas. There is nothing that makes it dissipate violently as it propagates. At HST-1, the jet suddenly bends by $\sim 16^{\circ}$. The bend seems to have disrupted the propagation, causing the spine to dissipate, making it conspicuous for the first time along its flow. Perhaps, an obstruction causes the jet deflection. There is the HST detection of an ionized disk of gas 0".25 from the nucleus \citep{for94}. There are density enhancements near the jet axis, so an obstruction is certainly plausible.

\par In order to assess the estimated value of $Q_{\rm{spine}}$, it is useful to provide the context of the surrounding tubular jet. In the region $0.35\, \rm{mas}<z< 0.65 \,\rm{mas}$, in 2013 and 2014, it is a mildly relativistic, protonic tubular region that comprises $\approx 58\%$ of the total jet volume \citep{pun23}. Each arm of the bilaterally symmetric tubular jet of radius, $R$, and wall thickness, $W$, transports $Q_{\rm{tubular\,jet}}\approx [(W/0.25R)^{0.46}]5.3\times 10^{41} \rm{ergs/s}$ \citep{pun22}. The wall thickness of the jet in this region was estimated to be $W\approx 0.35R$ \citep{pun23}. This implies a tubular jet power of
$Q_{\rm{tubular\,jet}}\approx 6.1\times 10^{41} \rm{ergs/s}$.

\par The total jet power (assuming bilateral symmetry) in the M87 frame is $Q(\rm{M87})\equiv 2[Q_{\rm{spine}}+Q_{\rm{tubular\,jet}}]$. Combining the spine and tubular jet power estimates is complicated by the different epochs of ejection from the central engine. The emission time of the spinal plasma at HST-1, $\sim 870\rm{mas}/(4.5\rm{mas/yr}) \approx 200$ years prior to the observation, is estimated using a speed of $\approx 4.5$ mas/yr from \citep{che07}. This is a crude estimate for demonstrative purposes only, all we know is a relativistic velocity is expected in the spine and $\beta_{\rm{app}}=1.14\pm 0.14$ (4.5 mas/yr) is consistent with this. An ejection time of 1.5 years before the observation of the tubular jet on sub-mas scales was estimated in Figure 2 of \citep{pun21}. New high resolution VLBI images indicate that the tubular jet emerges from the central engine thick-walled, i.e., it is not a Kelvin-Helmholtz instability generated boundary layer of the spine created farther downstream \citep{lu23}. If one assumes that the fundamental physical process in the central engine that launches the two component jet is the same for 200 years then it reasonable to assume that it would preferentially channel most of its jet power into either the spine or tubular jet for 200 years. If the spine (tubular jet) is more powerful, $Q(\rm{M87})$ emitted from the central engine was $Q(\rm{M87})<4Q_{\rm{spine}}\approx 1.0 \times 10^{42}\rm{ergs/s}$ ($Q(\rm{M87})< 4Q_{\rm{tubular\,jet}}\approx 2.4\times 10^{42}\rm{ergs/s}$) $\sim 200$ ($\sim1.5$) years before the observation.   The weakest conclusion that one can assert is $Q (\rm{M87}) < 2.4\times 10^{42} \rm{ergs/s}$ at some instance in time in the last $\sim 200$ years, provided that the equipartition assumption of the \citet{har06} models is not grossly inaccurate (the tubular jet power estimate does not rely on this assumption). Alternatively, assuming a nearly constant central engine injection jet power for $\sim 200$ years indicates a total jet power of $Q(\rm{M87})\lesssim 2\times 10^{42}$ ergs/s (i.e., typical of a Fanaroff-Riley 1 radio galaxy) in epochs of modern observation. This analysis indicates that the spine is not a powerful hidden reservoir for jet energy for the last 200 years. $Q(\rm{M87})$ found here is a factor of 30-300 less than estimates based on features ejected from the nucleus many hundreds to millions of years earlier \citep{owe00,for05,deg12}.

\par The bolometric luminosity, $L_{bol}$, of the inner jet is a consistency check. Before the EHT image, the inner jet emission was inseparable from disk emission in $L_{bol}$ estimates \citep{pri16}. Subtracting the EHT millimeter disk flux density in \citet{aki19} from lower resolution, quasi-simultaneous broadband data indicates $L_{\rm{bol}}(\rm{observed})\approx 5-6 \times 10^{41}$ erg/s for $z< 0.4$ arcseconds, even during a flare state (Punsly 2023 in preparation). From Equations (1) and (2), $L_{bol}$ in the M87 frame will be Doppler de-boosted. However, the region producing the peak of the SED is unresolved and the relevant $\beta$ (and de-boosting) is unknown. Regardless, $Q\sim 2\times10^{42}\rm{ergs/sec}$ is a sufficient energy budget to support $L_{bol}$.

\par  The jet efficiency, $\eta_{\rm{jet}}$, is defined by $Q(\rm{M87})\lesssim 2\times 10^{42} \rm{ergs/sec} =\eta_{\rm{jet}}\dot{M}c^{2}$, where $\dot{M}$ is the accretion rate, $\eta_{\rm{jet}}\lesssim0.035[(0.001M_{\odot}/\rm{yr})/\dot{M}]$. $\dot{M}$ must be larger than the mass flow rate of the sub-mas tubular jet, $\approx 0.00014(W/0.35R)^{0.46}M_{\odot}/\rm{yr}$ \citep{pun22}. For example, the EHTC has estimated ${\dot{M}}\approx 2.7\times 10^{-3}M_{\odot}/\rm{yr}$ in the single zone approximation \citep{aki20}. Alternatively, if most of the accreted mass is ejected in the jet, $\eta_{\rm{jet}} \lesssim 0.25$. Apparently, there is no requirement of black hole spin as a power source unless all the accreted mass is ejected in the jet in which case it is likely needed. The requirement that the jet needs a black hole spin power source might be an artifact of comparing jet powers in the distant past with current nuclear luminosity.

\begin{acknowledgements}
Many thanks to Yongjun Chen for the VLA image FITS files and C.C. Cheung for the VLBA image FITS file and Gaussian fit. This manuscript benefitted from the improvements suggested by a supportive referee.
\end{acknowledgements}

%

\begin{thebibliography}{}
\bibitem[Abramowski et al.(2012)]{abr12}Abramowski, A., Acero, F., Aharonian, F., et al. 2012, ApJ, 746, 151
\bibitem[Asada et al.(2016)]{asa16}Asada, K., Nakamura, M., Pu, H-Y., 2016, ApJ, 833, 56
\bibitem[Biretta et al.(1999)]{bir99}Biretta, J. A., Sparks, W. B., \& Macchetto, F. 1999, ApJ, 520, 621
\bibitem[Chen et al.(2011)]{che11}Chen, Y. J., Zhao, G.-Y., \& Shen, Z.-Q. 2011, MNRAS, 416, L109
\bibitem[Cheung et al.(2007)]{che07}Cheung, C. C., Harris, D. E., \& Stawarz, Ł. 2007, ApJL, 663, L65
\bibitem[Condon et al.(1998)]{con98}Cotton, J., Cotton, W., Greisen, E. et al. 1998, AJ, 115 1693
\bibitem[de Gasperin et al.(2012)]{deg12}de Gasperin, F., Orrú, E., Murgia, M., et al. 2012, A\&A, 547, 56
\bibitem[EHT Collaboration et al.(2019b)]{aki20} Event Horizon Telescope Collaboration, Akiyama, K., Alberdi, A., et al. 2019, ApJL, 875, L5
\bibitem[EHT Collaboration et al.(2019a)]{aki19} Event Horizon Telescope Collaboration, Akiyama, K., Alberdi, A., et al. 2019, ApJL, 875, L4
\bibitem[Ford et al.(1994)]{for94}Ford, H., Harms, R., Tsvetanov, Z. et al. 1994, ApJ, 435, L27
\bibitem[Forman et al.(2005)]{for05}Forman, W., Nulsen, P., Heinz, S., et al. 2005, ApJ, 635, 894
\bibitem[Gabuzda(2018)]{gab19}Gabuzda, D.2018 Galaxies 6 9
\bibitem[Gabuzda et al.(2018)]{gab18}Gabuzda, D.; Nagle, M.; Roche, N. 2018 A\& A 612 67
\bibitem[Ghisellini et al.(2005)]{ghi05}Ghisellini, G., Tavecchio, F., \& Chiaberge, M. 2005 A\& A 432 401
\bibitem[Ginzburg and Syrovatskii(1969)]{gin69} Ginzburg, V. and Syrovatskii, S. 1969, \araa, 7 375
\bibitem[Giroletti et al.(2012)]{gir12} Giroletti, M., Hada, K., Giovannini, G., et al. 2012, A\&A, 538, L10
\bibitem[Hada et al.(2016)]{had16} Hada, K., Kino, M., Doi, A., et al. 2016, ApJ, 817, 131
\bibitem[Hada(2017)]{had17} Hada, K. 2017, Galaxies, 5, 2; doi:10.3390/galaxies5010002
\bibitem[Harris et al.(2003)]{har03}Harris, D. E., Biretta, J. A., Junor, W., Perlman, E. S., Sparks, W. B., \& Wilson, A. S. 2003, ApJ, 586, L41
\bibitem[Harris et al.(2006)]{har06} Harris, D. E., Cheung, C. C., Biretta, J. A., et al. 2006, ApJ, 640, 211
\bibitem[Harris et al.(2009)]{har09} Harris, D. E., Cheung, C. C., Stawarz, Ł., Biretta, J. A., \& Perlman, E. S. 2009, ApJ, 699, 305
\bibitem[Lu et al.(2023)]{lu23} Lu, RS., Asada, K., Krichbaum, T.P. et al. A ring-like accretion structure in M87 connecting its black hole and jet. Nature 616, 686–690 (2023). https://doi.org/10.1038/s41586-023-05843-w
\bibitem[Medeiros et al.(2023)]{med23}Medeiros, L., Psaltis, D., Lauer, T., Ozel, F. 2023, ApJL, 947, 7
\bibitem[Owen et al.(2000)]{owe00}Owen F. N., Eilek J. A., Kassim N. E., 2000, ApJ, 543, 611
\bibitem[Perlman et al.(2011)]{per11}Perlman, E. S., Adams, S. C., Cara, M., et al. 2011, ApJ, 743, 119
\bibitem[Prieto et al.(2016)]{pri16} Prieto, M. A.; Fernandez-Ontiveros, J. A. Markoff, S. Espada, D. Gonzalez-Martin, O.  2016
MNRAS 457 3801
\bibitem[Porth et al.(2019)]{por19}Porth, O., Chaterjee, K., Narayan, R. et al. 2019 ApJS 243 26
\bibitem[Punsly(2008)]{pun08} Punsly, B. 2008, \emph{Black Hole Gravitohydromagnetics}, second edition (Springer-Verlag, New York)
\bibitem[Punsly(2021)]{pun21} Punsly, B. 2021, ApJ 918, 4
\bibitem[Punsly and Chen(2021)]{pun22}Punsly, B. and Chen, S. 2021, ApJL 921 L38
\bibitem[Punsly(2022)]{pun23}Punsly, B. 2022, ApJ 936, 79
\bibitem[Rampadarath et al.(2009)]{ram09}Rampadarath, H.; Garrett, M. A.; Polatidis, A. 2009, A \& A, 500, 1327
\bibitem[Reed(1989)]{ree89} Reed, B. 1989, Am. J. Phys., 57, 642
\bibitem[Rees(1966)]{ree66} Rees, M. J. 1966, \nat, 211, 468
\bibitem[Ro et al.(2023)]{ro23} Ro, H., Yi, K. Cui, Y. et al. 2023, Galaxies, 11, 33
\bibitem[Stawartz et al.(2006)]{sta06} Stawarz, Ł., Aharonian, F., Kataoka, J., et al. 2006, MNRAS, 370, 981
\bibitem[Tavecchio and Ghisellini (2008)]{tav08}Tavecchio, F. \& Ghisellini, G. 2008 MNRAS 385 L98
\bibitem[Walker et al.(2018)]{wal18} Walker R. C., Hardee P., Davies, F., Ly, C., Junor, W., 2018, ApJ 855 128
\end{thebibliography}
%

\end{document}